\documentclass[a4paper,11pt]{article}
\pdfoutput=1 

\usepackage{jheppub} 

\usepackage[T1]{fontenc} 

\usepackage{amssymb}
\usepackage{amsthm}
\usepackage{bookmark}
\usepackage{tabstackengine}

\allowdisplaybreaks

\title{\boldmath $\epsilon$-regular basis for non-polylogarithmic multiloop integrals and total cross section of the process $e^+e^-\to 2(Q\bar Q)$.}

\author[a]{ R.N. Lee}
\author[b,c]{A.I. Onishchenko}

\affiliation[a]{Theory department, Budker Institute of Nuclear Physics,\\ Novosibirsk, Russia}
\affiliation[b]{Bogoliubov Laboratory of Theoretical Physics, Joint
	Institute for Nuclear Research,\\ Dubna, Russia}
\affiliation[c]{Skobeltsyn Institute of Nuclear Physics, Moscow State University\\ Moscow, Russia}

\emailAdd{r.n.lee@inp.nsk.su}
\emailAdd{onish@theor.jinr.ru}

\abstract{We argue that in many physical calculations where the ``elliptic'' sectors are involved, one can express the results via iterated integrals with almost all weights being rational. Our method is based on the existence of $\epsilon$-regular basis, which is akin to the $\epsilon$-finite basis defined in Ref. \cite{Chetyrkin:2006dh}. As a demonstration of our technique, we calculate the photon contribution to the total cross section of the production of two $Q\bar Q$ pairs in the electron-positron collisions.
}

\definecolor{gray}{rgb}{0.800781, 0.800781, 0.800781}

\begin{document} 
\maketitle
\flushbottom

\section{Introduction}
Multiloop calculations have been an extremely rapidly developing subject in quantum field theory for the last thirty years. The two important milestones in the beginning of this long journey have been the invention of the integration-by-part reduction \cite{ChetTka1981,Tkachov1981} and differential equations technique \cite{Kotikov1991b,Remiddi1997}. These achievements have paved the way for the computer-assisted calculations which resulted in a burst-like growth of many physically relevant calculations. While doing these calculations, we have learned that the results can often be expressed in terms of iterated integrals with rational weights, viz., the multiple polylogarithms (MPLs), \cite{Goncharov1998,Remiddi:1999ew}. This class of functions is thoroughly investigated in many aspects, including the identities among them and algorithms for their effective numerical evaluation.

However, for some, at first relatively rare, cases it appeared that other functions may also be involved, the (arguably) simplest example being the phase-space integral for three massive particles. Later it was realized that the possibility to express the results in terms of MPLs corresponds\footnote{with some reservations related to the possibility to find a rationalizing variable change.} to the existence of $\epsilon$-form of the differential system \cite{Henn2013,Lee2014}. The transformation to this form resembles passing to the interaction picture in quantum mechanics, with $\epsilon=2-d/2$ being the small parameter, and the iterated integrals arising naturally from the perturbative expansion of a path-ordered exponent (evolution operator). In Ref. \cite{Lee2017c} a criterion of reducibility has been given. The non-polylogarithmic loop integrals correspond to the case where one can not express the ``unperturbed'' evolution operator in terms of rational or algebraic functions. In particular, for 3- and 4- equal-mass particles phase space integrals, this operator involves the complete elliptic integrals \cite{BAUBERGER1995383,Primo:2017ipr}. There is probably not much we can do about this fact, but the price we have to pay multiplies when we go to higher orders in $\epsilon$. Namely, each successive integration involves the kernel which contains functions entering the unperturbed evolution operator. The situation  only gets more complicated when these results enter the right-hand side of the differential equations for the master integrals in higher sectors. This is, of course, a poor man's approach, which, for the above-mentioned examples, is superseded by two alternative approaches, one resulting in the iterative integrals over modular forms \cite{Adams:2015ydq,Adams:2017ejb}, the other giving up the iterative structure of the integrals in favor of algebraic weights \cite{Broedel:2017siw,Broedel:2018qkq}\footnote{Here by ``iterated integral'' we mean the integral of the form $\int_{x_0<x_1<\ldots <x_n<x }dx_1 f_1(x_1)\ldots dx_n f_n(x_n)$, where the dependence on the kinematic parameter is only via the upper limit $x$.} (see also Refs. \cite{Remiddi:2016gno,Remiddi:2017har}). While we readily acknowledge a nice geometric picture behind each of these two approaches, one might nevertheless wonder if there is a narrower, simpler class of functions sufficient for the physical applications. A hint for the positive answer to this question can be found already  in the paper by G. Racah \cite{Racah1934a} written in 1934! In this paper the total cross section of the $e^+e^-$-pair production by a photon with energy $\omega$ in the field of a nucleus with charge $Z|e|$ has been calculated. The result was obtained by the direct integration of the spectrum and has the following form\footnote{A typographical error in the numerical coefficient of the last term has been corrected in Ref. \cite{Racah1936} by Racah himself. Note also that we use modern convention for the arguments of elliptic integrals, so that, e.g., $\mathrm{K}(z)=\intop_0^{\pi/2}\frac{d\theta}{\sqrt{1-z\sin^2\theta}}.$}
\begin{align}\label{eq:Racah}
	\sigma(\gamma Z\to e^+e^- Z)&=\frac{\alpha(Z\alpha)^2}{m^2}\bigg\{
	\frac{692+468\xi+76\xi^2+108\xi^3}{27(1+\xi)^3} \mathrm{K}(\xi^2)-\frac{692+360\xi+692\xi^2}{27(1+\xi)^3} \mathrm{E}(\xi^2)\nonumber\\
	&-\frac{4(1-\xi)^2}{(1+\xi)^2}\int_0^{\xi}\frac{\mathrm{K}(\eta^2)d\eta}{1-\eta}
	+\frac{16(1-\xi)^2}{(1+\xi)^2}\int_0^{\xi}\frac{d\zeta}{1-\zeta^2}\int_0^{\zeta}\frac{\mathrm{K}(\eta^2)d\eta}{1-\eta}
	\bigg\}\,,
\end{align}
where $\mathrm{K}$ and $\mathrm{E}$ are the complete elliptic integrals of the first and second kinds, respectively, and $\xi=\frac{\omega-2m}{\omega+2m}$ ($m$ is the electron mass). Note the modern look of this result which at the Racah time could have been underestimated. The appearance of the elliptic integrals is not surprising: they come due to the cut sunrise integral with two massive lines and one HQET line. However the rational integration weights are very remarkable. On general ground we would have expected the proliferation of transcendental weights in each successive integration. In the present paper we demonstrate that, in some sense, this is a general pattern. Our method is based on the existence of a special basis of master integrals which we call $\epsilon$-regular basis. It has many features common to $\epsilon$-finite basis  of Ref. \cite{Chetyrkin:2006dh}. Speaking loosely, our $\epsilon$-regular basis is that which remains a well-defined basis at $\epsilon=0$.

As an immediate application of our method, we calculate the photon contribution to the total Born cross section of the process $e^+e^-\to 2(Q\bar Q)$ with the full account of the quark mass. 

\section{\texorpdfstring{$\epsilon$}{Epsilon}-regular basis}\label{sec:epr}

Let us elaborate on the notion of $\epsilon$-regular basis of master integrals. While our definition is akin to that of $\epsilon$-finite basis in Ref. \cite{Chetyrkin:2006dh}, it somehow differs in a few details.

Suppose that we have a family $\mathcal{F}=\{j(k|\boldsymbol{n})|k=1,\ldots, K\text{ and } \boldsymbol{n}\in \mathbb{Z}^{N_k}\}$ of $L$-loop integrals which may belong to one or more prototypes (big topologies):
\begin{equation}
	j(k|n_1\ldots n_{N_k})=\int \frac{d^dl_1\ldots d^dl_L}{D_{k1}^{n_1}\ldots D_{k N_k}^{n_{N_k}}}\,.
\end{equation} 
Here $d=4-2\epsilon$ and $k=1,\ldots, K$ enumerates different prototypes of our family.
We define the \textit{basis} of $\mathcal{F}$ as a finite set of functions $j_1,\ldots j_M$ such that any integral of $\mathcal{F}$ can be uniquely represented as their linear combination with coefficients being the rational functions of kinematic invariants and space-time dimension over rational numbers. In what follows we will always understand the linear combinations and linear (in)dependence in this way. In particular, from the uniqueness we conclude that $j_1,\ldots j_M$ are all linearly independent. This is a usual definition of the basis of master integrals except that we allow also for the linear combinations of integrals as its elements (as ``master integrals''). Then it is easy to understand that there exists such a basis $F_1,\ldots F_M$ that the following conditions are fulfilled:
\begin{enumerate}
	\item The $\epsilon$-expansion of each $F_m$ starts from $\epsilon^0$, i.e., $F_m=F_m^{(0)}+O(\epsilon)$. 
	\item The leading terms $F_1^{(0)},\ldots, F_M^{(0)}$ are linearly independent (in the above sense).
\end{enumerate}
We will call any basis with these properties the \textit{$\epsilon$-regular basis}.
From the second condition it follows, in particular, that  $F_m^{(0)}\neq 0$ for any $m=1,\ldots, M$.

Let us first explain why such a basis necessarily exists. Suppose that we have arbitrary basis $F_1,\ldots, F_M$, then we can multiply each element by a suitable power of $\epsilon$ so that the redefined basis satisfies the first condition. Now assume $F_1,\ldots, F_M$ satisfies the first condition, but there is a vanishing linear combination  $F_{m_0}^{(0)}-\sum_{m<m_0}c_m F_m^{(0)}=0$. Then we redefine the $m_{0}$-th element of the basis by replacing
\begin{equation}
	F_{m_0}\to \frac{1}{\epsilon}[F_{m_0}-\sum_{m<m_0}c_m F_m]\,.
\end{equation} 
It is easy to understand that proceeding in this way we will finally obtain the $\epsilon$-regular basis\footnote{Here we silently rely on some properties of the multiloop integrals. In particular, we assume that consecutive terms of $\epsilon$ expansion generate infinite or, at least, large enough set of mutually transcendental functions to form the suitable basis. For generic set, the $\epsilon$-regular basis does not necessarily exists. Consider, e.g., the linear span of $F_1(x)=\exp(\epsilon x),F_2(x)=\exp(-\epsilon x)$. Obviously, the described process will not terminate for these two functions.}. 

The rationale behind our definition is the following. Suppose that we want to calculate some quantity $P$ whose $\epsilon$-expansion starts from $O(\epsilon^0)$ and which is a linear combination of the integrals of $\mathcal{F}$ (possibly, with coefficients singular at $\epsilon=0$). Then this quantity can be reduced to linear combination of elements $F_1,\ldots,F_M$ of $\epsilon$-regular basis with coefficients which are necessarily regular at $\epsilon=0$. The proof by contradiction is almost trivial: suppose there are singular coefficients with the maximal pole order $n>0$. Then the coefficient of $\epsilon^{-n}$ is a non-trivial linear combination of the leading terms $F_1^{(0)},\ldots,F_M^{(0)}$, which, by second condition, can not be zero. Therefore, the expansion of $P$ should start from $\epsilon^{-n}$, which contradicts our assumption. Therefore, we can claim that any physical quantity well-defined at $\epsilon=0$ and expressed, in dimensional regularization, as a linear combination of multiloop integrals of $\mathcal {F}$, is a linear combination of the leading terms $\mathcal{F}_1^{(0)},\ldots,\mathcal{F}_M^{(0)}$ of $\epsilon$-regular basis. There is another important property of the $\epsilon$-regular basis. Consider the differential system $d\boldsymbol{F}=M_F \boldsymbol{F}$. Then we can claim that $M_F$ has a finite $\epsilon\to 0$ limit. The proof goes along the same lines as above.

Let us now assume that the differential system for some master integrals $\boldsymbol{J}=(J_1,\ldots, J_M)^\intercal$ is in $\epsilon$-form:
\begin{equation}
	d\boldsymbol{J}=\epsilon A(\boldsymbol{x}) \boldsymbol{J}\,,
\end{equation}
where $\boldsymbol{x}$ denotes the kinematic parameters, and the matrix 1-form $A$ satisfies $dA=A\wedge A=0$.
Then there is an $\epsilon$-regular basis $\boldsymbol{F}=(F_1,\ldots, F_M)^\intercal$  satisfying
\begin{equation}\label{eq:FBDE}
d\boldsymbol{F}=B(\epsilon,\boldsymbol{x}) \boldsymbol{F}=\sum_{k=0}^{K}\epsilon^kB_k(\boldsymbol{x}) \boldsymbol{F}\,,
\end{equation}
where $K<\infty$ and $B_0$ is strictly lower triangular, i.e., lower triangular with zero diagonal. In order to prove this, let us formulate two ``moves'' which will eventually render $\boldsymbol{J}$ into $\boldsymbol{F}$. Preliminary step is to multiply all $\boldsymbol{J}$ by a common factor $\epsilon^n$, where $n$ is chosen so that the modified basis $\widetilde{\boldsymbol{J}}=\epsilon^n\boldsymbol{J}$ is regular at $\epsilon=0$ and for some $m$ we had $\widetilde{\boldsymbol{J}}_m^{(0)}\neq0$. These requirements uniquely fix $n$. In what follows we will denote our current basis by $\boldsymbol{J}$.

From now on we repetitively do the following. On each repetition we assume that $B(\epsilon,\boldsymbol{x})=\sum_{k=0}^{K<\infty}\epsilon^kB_k(\boldsymbol{x})$ and that $B_0(\boldsymbol{x})$ is strictly lower triangular and check that these assumptions also hold in the end of the cycle. 
Assume that the first $m-1$ integrals satisfy both conditions, but first $m$ integrals don't. It means that either integral $J_{m}$ starts from higher order of $\epsilon$, or that its leading term $J_{m}^{(0)}$ is linearly dependent on $J_{1}^{(0)},\ldots,J_{m-1}^{(0)}$, i.e., that $J_{m}^{(0)}=c_1J_{1}^{(0)}+\ldots+c_{m-1}J_{m-1}^{(0)}$ with some rational coefficients $c_1,\ldots,c_{m-1}$. 
\begin{enumerate}
	\item Let $J_{m}$ start from higher order of $\epsilon$. We replace $J_m$ with $\tilde{J}_m=\epsilon^{-1}J_m$. After this replacement the matrix $B(\epsilon,\boldsymbol{x})$ in the right-hand side of the differential system is altered. Namely, the $m$-th row is divided by $\epsilon$ and the $m$-th column is multiplied by $\epsilon$, while $B_{m,m}$ is unchanged. Let us first  prove that $B(\epsilon,\boldsymbol{x})$ is regular at $\epsilon=0$. Obviously, the poles might appear only on $m$-th row in positions $1,\ldots,m-1$. However, that would mean that some linear combination of $J_1,\ldots, J_{m-1}$ is vanishing at $\epsilon\to 0$, which by assumption is not the case. So, there are no poles in the new matrix $B$. Some elements on $m$-th row, starting from position $m+1$ might now be of order $\epsilon^0$. Therefore, the new matrix $B_0$ is not strictly lower triangular one. However, this is fixed by re-numerating the integrals. Note that $m$-th column of the matrix $B$ is suppressed at least as $\epsilon$. 	Let the index of last nonzero element of the $m$-th row of new $B_0$ be $\tilde{m}$ ($\tilde{m}>m$).  Then we re-numerate integrals $J_{m+1},\ldots, J_{\tilde{m}}$ to be the new $\tilde{J}_{m},\ldots, \tilde{J}_{\tilde{m}-1}$ and $\tilde{J}_m$ to be the new $\tilde{\tilde{J}}_{\tilde{m}}$. So, the overall change is
	\begin{equation}
		\boldsymbol{J}=
		\begin{pmatrix}
		\vdots \\ J_{m}\\ J_{m+1}\\ \vdots\\ J_{\tilde{m}}\\ \vdots
	\end{pmatrix}
		\longrightarrow
	\begin{pmatrix}
		\vdots \\ J_{m+1}\\ \vdots\\ J_{\tilde{m}}\\\epsilon^{-1}J_{m}\\ \vdots
	\end{pmatrix}\,.
	\end{equation}
	After this change the new matrix $B$ is regular at $\epsilon=0$ and its leading term $B_0$ is strictly lower triangular.
	\item Let  $J_{m}^{(0)}=c_1J_{1}^{(0)}+\ldots+c_{m-1}J_{m-1}^{(0)}$. Then we pass from $J_{m}$ to $\tilde{J}_{m}=J_{m}-c_1J_{1}-\ldots-c_{m-1}J_{m-1}$. This new integral is suppressed at least as $\epsilon^1$, so on the next iteration we will hit case 1. 
	Thus, the change is
	\begin{equation}
	\boldsymbol{J}=
	\begin{pmatrix}
	\vdots\\ J_{m}\\ \vdots
	\end{pmatrix}
	\longrightarrow
	\begin{pmatrix}
	\vdots\\ J_{m}-\sum_{k=1}^{m-1}c_kJ_k\\ \vdots
	\end{pmatrix}\,.
	\end{equation}
	It is clear that the new matrix $B$ is regular at $\epsilon=0$ and $B_0$ is strictly lower triangular.
\end{enumerate}

Naturally, when we have processed the case $m=M$, the algorithm terminates, and the current set of master integrals is obviously an $\epsilon$-regular basis $\boldsymbol{F}$. Suppose that our initial matrix $A$ was lower block-triangular with diagonal blocks corresponding to sectors. Then the above algorithm tries to conserve the block structure as much as possible. 

From Eq. \eqref{eq:FBDE} we see that the leading coefficients satisfy the equation
\begin{equation}
d\boldsymbol{F}^{(0)}=B_0(\boldsymbol{x}) \boldsymbol{F}^{(0)}\,,
\end{equation}
which can be solved in quadratures as $\boldsymbol{F}^{(0)}=\int B_0(\boldsymbol{x}) \boldsymbol{F}^{(0)}$,
and, thanks to the strict lower-triangularity of the matrix $B_0$, this does not lead to circular definition.

Suppose now that some sectors can not be reduced to $\epsilon$-form. Then, proceeding in a similar way, we will still be able to obtain the same form  \eqref{eq:FBDE}, where however the matrix $B_0$ contains now some non-zero diagonal blocks. The solution of the corresponding homogeneous system is then nontrivial. However, once this solution is known, the leading terms of the integrals in the higher sectors are obtained as iterated integrals with rational weights.

\section{Pedagogical example: derivation of Racah result for $\sigma_{B}(\gamma Z\to e^+e^- Z)$.}

\begin{figure}
	\centering\includegraphics[width=0.7\linewidth]{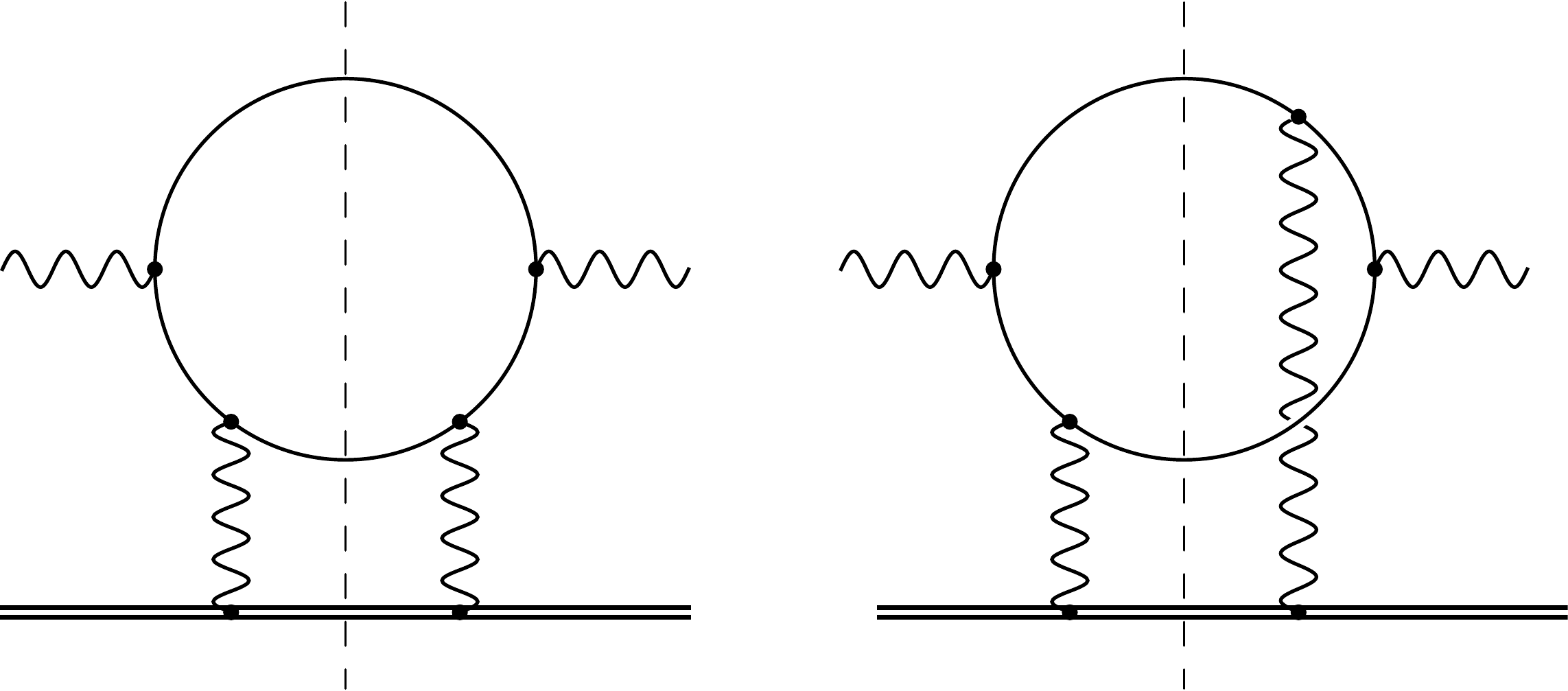}	
	\caption{Cut diagrams for the calculation of the total cross section of $e^+e^-$ pair production by a photon in the field of nucleus. Cut thin line denotes the cut propagators $-2\pi i\delta(p^2-m^2)(\hat{p}+m)$ of the electron, cut double line denotes the cut propagator $-2\pi i\delta(2 q\cdot v)=-2\pi i\delta(2 q_0)$ of a heavy particle, interaction vertex with the heavy particle is $-iv^\mu$ ($v=(1,\boldsymbol{0})$ is a four-velocity of the nucleus). }
	\label{fig:racah1}
\end{figure}

Let us first demonstrate how the Racah result is obtained via the modern differential equations technique. The total cross section of the process is expressed as a sum of diagrams depicted in Fig. \ref{fig:racah1}.

Contributions of both diagrams can be expressed via integrals of the following form
\begin{gather}
j_{n_1\ldots n_{7}}=\int \frac{d^dl\,d^dq}{(2\pi)^{2d}}\theta(q^0-l^0)\theta(l^0+k^0)\prod_{k=1}^3\Im\frac{1}{(D_k-i0)^{n_k}}\prod_{k=4}^{7}\frac{1}{(D_k-i0)^{n_k}}\,,\nonumber\\
D_1=1-(k+l)^2\,,\quad 
D_2=1-(l-q)^2\,,\quad
D_3=-2 q\cdot v=-2 q_0\,,\nonumber\\
D_4=1-l^2\,,\quad
D_5=1-(k+l-q)^2\,,\quad
D_6=-q^2\,,\quad 
D_7=-2 l\cdot v=-2 l_0\,,
\end{gather}
where $D_7$ denotes an irreducible numerator. Here and below we put the electron mass to unit for simplicity.

The IBP reduction\footnote{We use \texttt{LiteRed} for the IBP reduction, \cite{Lee2013a}.} reveals 4 master integrals:
\begin{equation}
	j_{1110000},j_{1120000},j_{1110010},j_{1110100}\,.
\end{equation}
It is easy to understand that all these integrals are finite at $\epsilon=0$. However, if we express the cross section via these master integrals, the coefficients will have poles:
\begin{multline}
	\sigma=\frac{(16\pi)^3\alpha (Z\alpha)^2}{\omega}\bigg\{
	\left(-\frac{1
	}{4 \epsilon }+\frac{7}{4}+O(\epsilon)\right)
	j_{1110010}
	+\left(\frac{1}{2\epsilon
	}-2+O(\epsilon)\right) j_{1110100}\\
	\left(\frac{3}{16 \epsilon }-\frac{69 \omega ^2+58}{72 \omega ^2}+O(\epsilon)\right)j_{1110000}
	+\frac{\omega^2 -4}{8\omega}\left(\frac{1}{\epsilon }-\frac{34 \omega ^2+4}{9 \omega ^2}+O(\epsilon)\right)j_{1120000} 
	\bigg\}\,,
\end{multline}
where $\omega=k_0$ and we have truncated the coefficients at $\epsilon^0$ to save space. Since we calculate the finite quantity, the $\epsilon^{-1}$ terms should cancel, which gives the condition
\begin{equation}\label{eq:constraint}
\frac12j_{1110100}^{(0)}-\frac{1}{4}j_{1110010}^{(0)}
+\frac{\omega^2 -4}{8\omega}j_{1120000}^{(0)}
+\frac{3}{16}j_{1110000}^{(0)}
 = 0\,,
\end{equation} 
where $j_{k}^{(0)}$ denotes the leading $\propto \epsilon^0$ term of $j_k$.

Before we proceed further, let us underline, that in usual approach we would have concluded here, that we need to know all integrals up to $\epsilon^1$. Meanwhile, the first two integrals can not be expressed via polylogarithms. Their leading terms $j_{1110000}^{(0)}$ and  $j_{1120000}^{(0)}$ read
\begin{align}
	j_{1110000}^{(0)}&=\frac{\left(\xi ^2+1\right) \mathrm{E}\left(\xi ^2\right)+\left(\xi ^2-1\right) \mathrm{K}\left(\xi ^2\right)}{24 \pi ^3 (1-\xi)^3}\,,\label{eq:j1}\\
	j_{1120000}^{(0)}&=-\frac{(\xi +1) \left(\mathrm{E}\left(\xi ^2\right)+(\xi -1) \mathrm{K}\left(\xi ^2\right)\right)}{64 \pi ^3 (\xi -1)^2}\,.\label{eq:j2}
\end{align}
Therefore, the next terms $j_{1110000}^{(1)}$ and  $j_{1120000}^{(1)}$ are likely to involve modular forms in the integration weights.

In the approach advocated here, we first pass to $\epsilon$-regular basis. In particular, thanks to the relation \eqref{eq:constraint}, we can pass from $j_{1110100}$ to the linear combination 
\begin{equation}
\tilde{j}=\frac{1}{2\epsilon}\left[
j_{1110100}-\frac{1}{2}j_{1110010}
+\frac{\omega^2 -4}{4\omega}j_{1120000}
+\frac{3}{8}j_{1110000}\right]\,.
\end{equation}
For the purpose of simpler presentation, we prefer to pass to a slightly different combination
\begin{equation}
	\frac{1-\epsilon }{2 \epsilon }j_{1110100}-\frac{1-3 \epsilon}{4 \epsilon } j_{1110010}+\frac{\omega^2 -4}{8 \omega  \epsilon}j_{1120000}+\frac{3-4 \epsilon }{16 \epsilon }j_{1110000}=j_{1110110}\,,
\end{equation}
whose coefficients differ from those in $\tilde{j}$ only by $\epsilon^0$ terms, but which can be represented as one specific integral $j_{1110110}$. In terms of new master integrals the cross section has the form
\begin{equation}
	\sigma=\frac{(16\pi)^3\alpha (Z\alpha)^2}{\omega}\bigg\{j_{1110110}+\frac{1}{4}j_{1110010}-\frac{(\omega^2 -4) (7 \omega ^2+4)}{72 \omega ^3}j_{1120000}-\frac{21 \omega ^2+116}{144 \omega ^2}j_{1110000}\bigg\}\,,
\end{equation}
where we have omitted $O(\epsilon^1)$ terms in the coefficients. Note the absence of poles in the coefficients. The differential system for the column
\begin{equation}
	\boldsymbol{j}=(j_{1110000},j_{1120000},j_{1110010},j_{1110110})^{\intercal}
\end{equation}
has the form $\partial_\xi \boldsymbol{j}=M\boldsymbol{j}$ with
\begin{equation}
	M=\left(
	\begin{array}{cccc}
	0 & -\frac{8}{(1-\xi)^2} & 0 & 0 \\
	\frac{(1-2 \epsilon) (3-4 \epsilon)}{8
		\xi } & \frac{(1-2 \epsilon)
		\left(\xi ^2+4 \xi
		+1\right)}{ \xi 
		(1-\xi^2)} & 0 & 0 \\
	0 & -\frac{2}{(1+\xi)^2} & -\frac{2 (1-2
		\epsilon)}{1-\xi^2} & 0 \\
	0 & -\frac{1}{2 (\xi +1)^2} & -\frac{1-3
		\epsilon}{1-\xi^2}
	& -\frac{2}{1-\xi^2}
	\\
	\end{array}
	\right)
\end{equation}
This system has a regular limit $\epsilon\to0$, therefore, from now on we can safely put $\epsilon=0$. As we have seen in the previous section, we can find the transformation which reduces the differential system to strictly lower-triangular form everywhere except the irreducible block corresponding to the ``elliptic'' sector. Indeed, passing to new master integrals $\boldsymbol{F}=(F_1,F_2,F_3,F_4)^\intercal$ related to $\boldsymbol{j}$ via
\begin{gather}\label{eq:jviaF}
\boldsymbol{j}=T\boldsymbol{F}\,,\\
T|_{\epsilon=0}=\left(
\begin{array}{cccc}
-\frac{16 \xi }{3(1-\xi^2)} & \frac{16
	(1+\xi ^2)}{3(1-\xi^2)} & 0 & 0 \\
0 & -2 & 0 & 0 \\
0 & \frac{1-\xi }{1+\xi} & -\frac{1-\xi }{1+\xi} & 0 \\
0 & 0 & 0 & \frac{1-\xi }{1+\xi} \\
\end{array}
\right)\,,
\end{gather}
we obtain the differential system
\begin{gather}\label{eq:deF}
	\partial_\xi \boldsymbol{F}=M_F\boldsymbol{F}\,,\\
	M_F|_{\epsilon=0}=\left(
	\begin{array}{cccc}
	0 & \frac{1-\xi}{\xi  (1+\xi )} & 0 & 0 \\
	\frac{1}{1-\xi ^2} & \frac{4}{1-\xi ^2} & 0 & 0 \\
	\frac{1}{1-\xi ^2} & 0 & 0 & 0 \\
	0 & 0 & \frac{1}{1-\xi ^2} & 0 \\
	\end{array}
	\right)
\end{gather}
Using Eqs. \eqref{eq:j1}, \eqref{eq:j2}, and \eqref{eq:jviaF} we obtain
\begin{align}
	F_1^{(0)}(\xi)&=\frac{1+\xi}{128 \pi ^3} \mathrm{K}\left(\xi ^2\right)\,,\nonumber\\
	F_2^{(0)}(\xi)&=\frac{1+\xi}{128 \pi ^3 (1-\xi)^2}\left[\mathrm{E}\left(\xi ^2\right)-(1-\xi) \mathrm{K}\left(\xi ^2\right)\right]
\end{align}
The results for $F_{3,4}^{(0)}$ can be obtained by direct integration of Eq. \eqref{eq:deF}:
\begin{align}
F_3^{(0)}(\xi)&=\intop_{0}^{\xi} d\eta \frac{F_1^{(0)}(\eta)}{1-\eta^2}=\frac1{128\pi^3}\intop_{0}^{\xi} \frac{\mathrm{K}(\eta^2) d\eta}{1-\eta}\,,\\
F_4^{(0)}(\xi)&=\intop_{0}^{\xi} d\zeta \frac{F_3^{(0)}(\zeta)}{1-\zeta^2}=\frac1{128\pi^3}\intop_{0}^{\xi} \frac{d\zeta}{1-\zeta^2}\intop_{0}^{\zeta} \frac{\mathrm{K}(\eta^2) d\eta}{1-\eta}\,.
\end{align}
The integration constants are fixed by the condition of vanishing of $F_{3,4}^{(0)}$ at the threshold.

In terms of the integrals $F_1,F_2,F_3,F_4$ the cross section has the form
\begin{multline}\label{eq:csvsF}
\sigma=\frac{(16\pi)^3\alpha (Z\alpha)^2}{m^2}\bigg\{\frac{\xi  \left(25 \xi ^2-8 \xi +25\right) F_1^{(0)}}{27 (1+\xi)^4}-\frac{(1-\xi)^2 \left(173 \xi ^2+90 \xi +173\right) F_2^{(0)}}{216 (1+\xi)^4}\\
-\frac{(1-\xi)^2 F_3^{(0)}}{8 (1+\xi )^2}+\frac{(1-\xi)^2 F_4^{(0)}}{2(1+\xi)^2}\bigg\}
\end{multline}
Substituting results for $F_{1-4}$ into Eq. \eqref{eq:csvsF}, we obtain the Racah result \eqref{eq:Racah}.

\section{Total Born cross section of $e^+e^-\to 2(Q\bar Q)$}

\begin{figure}
	\centering\includegraphics[width=1.\linewidth]{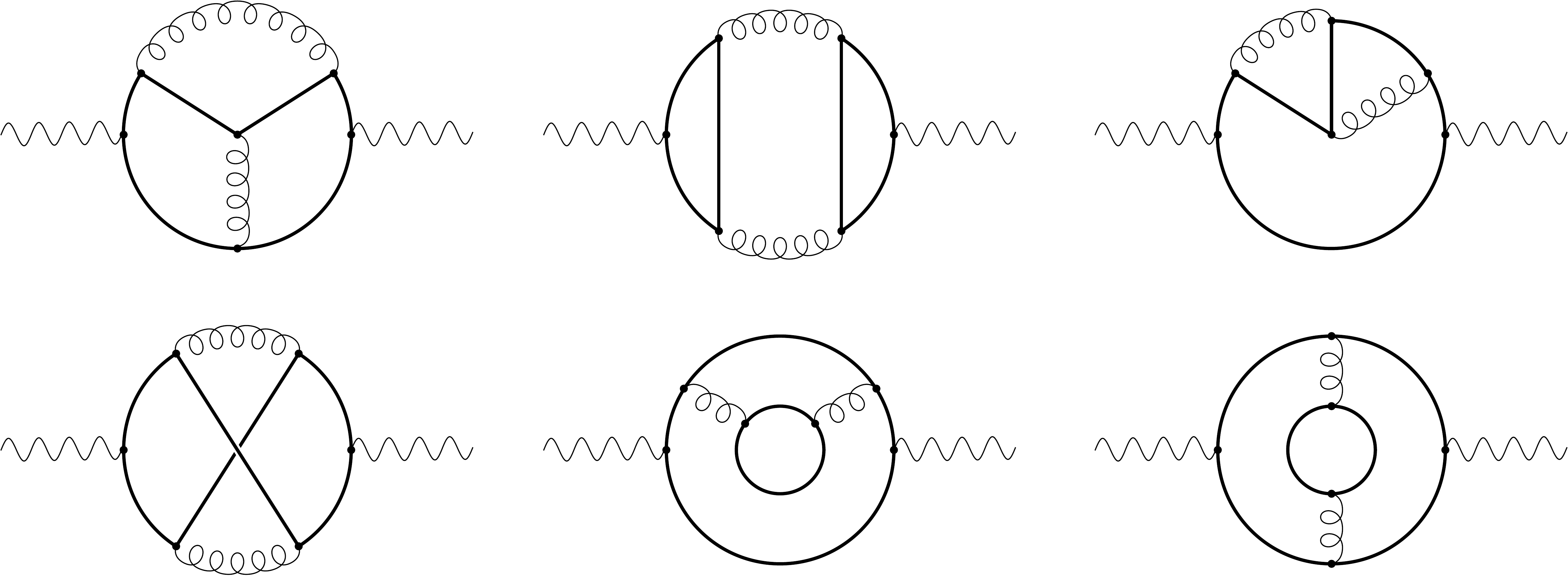}	
	\caption{Photon polarization diagrams with non-zero $4m$ cuts contributing to total cross-section of $e^+e^-$ annihilation into two heavy quark anti-quark pairs.}
	\label{fig:diags2e4Q}
\end{figure}

\begin{figure}
	\centering\includegraphics[width=0.9\linewidth]{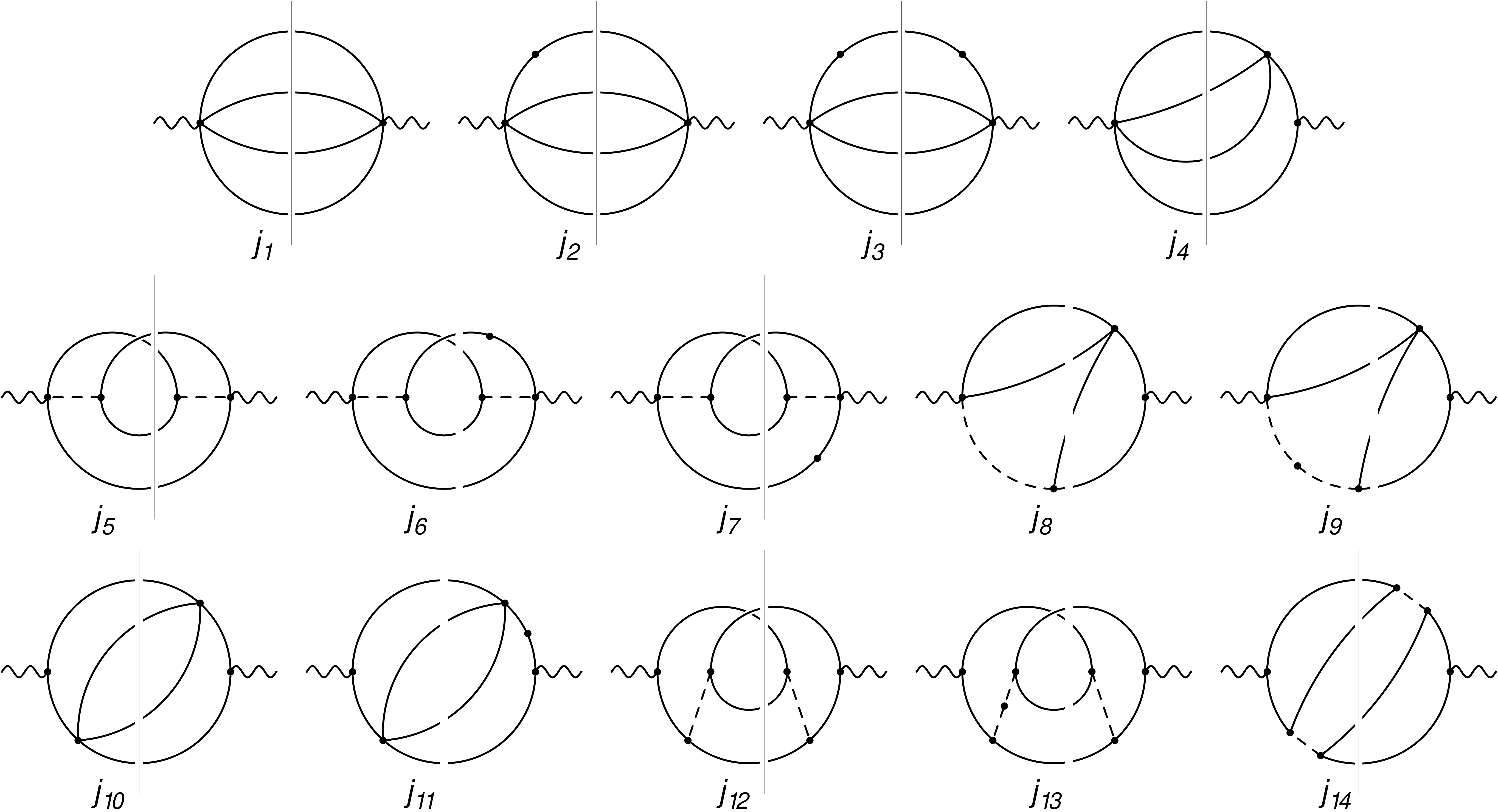}	
	\caption{The cut  master integrals. The solid (dashed) lines denote massive (massless) propagators, thin vertical line corresponds to the cut.}
	\label{fig:masters-4m-cut}
\end{figure}

Let us now apply the same technique to the calculation of total Born cross section for the production of two heavy quark anti-quark pairs in $e^+e^-$ annihilation\footnote{Here we restrict ourselves by $c$ or $b$-quarks pairs, when it is sufficient to consider only photon exchange contribution.}. The latter can be conveniently expressed via $4 m$ discontinuity of photon polarization operator as\footnote{By $4 m$ discontinuity we mean the contribution of cuts with four massive lines going on-shell.} 
\begin{equation}
\sigma = \frac{e^2}{6s^2}\text{Disc}_{4m} \Pi^{\mu}_{\mu}\, 
\end{equation}
where $e$ is electron charge and $\Pi^{\mu\nu}$ is the photon polarization operator. The corresponding diagrams are shown in Fig. \ref{fig:diags2e4Q}.
To calculate the $4m$ discontinuity of photon polarization operator we first use IBP relations to reduce the diagrams in Fig. \ref{fig:diags2e4Q} to master integrals and consider all $4m$ cuts of the latter. There are 14 distinct cut master integrals depicted in Fig. \ref{fig:masters-4m-cut}. 

The master integrals in the lowest sector are related to the phase space of four massive particles. They satisfy a homogeneous differential system whose solution  at $d=2$ in terms of the complete elliptic integrals was found in Ref. \cite{Primo:2017ipr}. Using the threshold asymptotics to fix the boundary conditions we have 
\begin{align}\label{eq:j1[d=2]}
	j_1|_{d=2}&=f(s)=\frac{16 \pi}{s}\left[\mathrm{K}\left(1-k_-\right) \mathrm{K}\left(k_+\right)-\mathrm{K}\left(k_-\right) \mathrm{K}\left(1-k_+\right)\right]\,,
\end{align}
where 
\begin{equation}
	k_{\pm}=\frac12\left[1\pm\left(1-\frac8s\right) \sqrt{1-\frac{16}{s}}+\frac{16}{s} \sqrt{1-\frac{4}{s}}\right]\,.
\end{equation}
At $d=4$, the master integrals $j_1|_{d=4},\,j_2|_{d=4},\,j_3|_{d=4}$ are expressed as linear combinations of $f(s)$, $f'(s)$ and $f''(s)$. The explicit form of these expressions is not essential at the moment. Therefore, if we write $j_k|_{d=4-2\epsilon}=\sum_n \epsilon^n j_k^{(n)}$ ($k=1,2,3$), we do know the leading coefficients $j_k^{(0)}$, but, to the best of our knowledge, not more. Meanwhile, if we express the total cross section via $j_{1-14}$, the coefficients in front of $j_{1,2,3,4}$ will have second order poles in $\epsilon$:
\begin{multline}
\sigma = \frac{\alpha^2\alpha_s^2e_Q^2}{24\pi^2s^4(s-4)\epsilon^2}
(17s^4+59s^3-146s^2-872s+576) \\
\times\left\{
\frac{1}{8}(8-s)j_1 + \frac{1}{3}(6-s)j_2 + \frac{1}{24}s(s-16)j_3 + \frac{1}{6}(4-s)j_4
\right\} + \mathcal{O}(\epsilon^{-1})\, .
\end{multline}
Therefore, within the standard approach we would have to calculate $j_k^{(n)}$ with $n=1,2$. This looks like a very challenging task. However, using the methods described in Section \ref{sec:epr}, we pass to $\epsilon$-regular basis $\boldsymbol{F}=(F_1,\ldots F_{14})^\intercal$,
\begin{equation}
	\boldsymbol{j} = T\boldsymbol{F}\,,
\end{equation}
where the matrix $T$ can be found in the ancillary file \texttt{Tj2F.m}. In terms of these new functions, the cross section reads
\begin{multline}
\sigma = \frac{\alpha^2\alpha_s^2e_Q^2}{27\pi^2 s^2}\Bigg\{
\frac{3 s^2+430 s+128}{96 s} F_1^{(0)}(s)
+\frac{85 s^3+990 s^2+16680 s-92608}{144 (s-4) s} F_2^{(0)}(s)\\
-\frac{1019 s^2-1286 s-12120}{36 s} F_3^{(0)}(s)
-\frac{(s-6) \left(49 s^2-336 s+464\right) \beta }{12(s-4)^2} F_4^{(0)}(s)\\
-\frac{(s-1) (s+2)}{s}F_5^{(0)}(s)
-\frac{2\left(s^2+2 s-18\right) \beta}{s-4}F_6^{(0)}(s)
+\frac{5 s^2-12 s-50}{5 s}F_7^{(0)}(s)\\
+\frac{5 s^2-2 s-60}{s-4}\beta F_8^{(0)}(s)
+\frac{s^2+s+4 }{s}F_9^{(0)}(s)
+2\frac{s^2-2}{s}F_{10}^{(0)}(s)
-\frac{2  (s+2) (s-2)^2\beta}{(s-4) s}F_{12}^{(0)}(s) 
\Bigg\}\, ,
\end{multline} 
where $e_Q$ is the heavy quark charge and $\beta = \sqrt{1-4/s}$. Note that only leading terms $F^{(0)}_k$ enter this expression, as it should be. The system of differential equations for the column $\boldsymbol{F}^{(0)}=(F_1^{(0)},\ldots, F_{14}^{(0)})^\intercal$ has the form $\partial_s \boldsymbol{F}^{(0)}= M_F \boldsymbol{F}^{(0)}$, where 
\begin{equation}
M_F= \resizebox{0.7\hsize}{!}{$\left(
\begin{array}{cccccccccccccc}
\frac{3}{2 (s-16)} & -\frac{1}{s}  &  \mspace{20mu}0\mspace{20mu} &  \mspace{20mu}0\mspace{20mu} &  \mspace{20mu}0\mspace{20mu} &  \mspace{20mu}0\mspace{20mu} &  \mspace{20mu}0\mspace{20mu} &  \mspace{20mu}0\mspace{20mu} &  \mspace{20mu}0\mspace{20mu} &  \mspace{20mu}0\mspace{20mu} &  \mspace{20mu}0\mspace{20mu} &  \mspace{20mu}0\mspace{20mu} &  \mspace{20mu}0\mspace{20mu} &  \mspace{20mu}0\mspace{20mu} \\
\frac{3}{2 (s-16)} & -\frac{1}{s} & -\frac{2}{s} & 0 & 0 & 0 & 0 & 0 & 0 & 0 & 0 & 0 & 0 & 0\\
-\frac{3}{16 s \beta ^2} & \frac{s-16}{8 s^2 \beta ^2} & -\frac{s+8}{2 s^2 \beta ^2} & 0 & 0 & 0 & 0 & 0 & 0 & 0 & 0 & 0 & 0 & 0 \\
\frac{3 \beta }{2 (s-16)} & -\frac{1}{s \beta } & 0 & 0 & 0 & 0 & 0 & 0 & 0 & 0 & 0 & 0 & 0 & 0 \\
\frac{1}{s} & 0 & 0 & 0 & 0 & 0 & 0 & 0 & 0 & 0 & 0 & 0 & 0 & 0 \\
0 & 0 & 0 & -\frac{1}{s \beta ^2} & 0 & 0 & 0 & 0 & 0 & 0 & 0 & 0 & 0 & 0 \\
0 & 0 & 0 & \frac{1}{s \beta } & 0 & 0 & 0 & 0 & 0 & 0 & 0 & 0 & 0 & 0 \\
\frac{1}{2 s \beta } & 0 & 0 & -\frac{1}{s} & 0 & 0 & 0 & 0 & 0 & 0 & 0 & 0 & 0 & 0 \\
0 & 0 & 0 & 0 & 0 & \frac{1}{s \beta } & 0 & -\frac{1}{s \beta } & 0 & 0 & 0 & 0 & 0 & 0 \\
0 & 0 & 0 & 0 & 0 & 0 & \frac{1}{s} & 0 & 0 & 0 & 0 & 0 & 0 & 0 \\
0 & 0 & 0 & 0 & 0 & \frac{1}{s^2 \beta ^2} & -\frac{3}{4 s \beta } & -\frac{4}{s^2 \beta ^2} & 0 & 0 & 0 & 0 & 0 & 0 \\
0 & 0 & 0 & 0 & -\frac{1}{4 s \beta } & 0 & \frac{1}{s \beta } & 0 & 0 & 0 & 0 & 0 & 0 & 0 \\
0 & 0 & 0 & 0 & 0 & 0 & 0 & 0 & 0 & 0 & \frac{1}{s \beta } & \frac{1}{s \beta } & 0 & 0 \\
0 & 0 & 0 & 0 & 0 & 0 & 0 & 0 & \frac{1}{s} & 0 & 0 & \frac{1}{s \beta } & 0 & 0 
\end{array}
\right)$}
\end{equation}
Apart from the $3\times3$ upper left block, the matrix $M_F$ is strictly lower triangular. Therefore, the integrals $F_{4,\ldots,14}$ can be expressed via iterated integrals of $F_{1-3}$. If we pass from $s$ to $\beta$ via $s=\frac{4}{1-\beta^2}$, it is easy to see that the weights in these iterated integrals are rational in $\beta$. Using the above matrix, it is trivial to write $F_4^{(0)},\ldots, F_{14}^{(0)}$ as iterated integrals of $F_1^{(0)}$ and $F_{2}^{(0)}$. It worth noting that all appearing iterated integrals can be transformed into one-fold integrals of MPLs and complete elliptic integrals. Explicit form of $F_1^{(0)},\ldots, F_{14}^{(0)}$ is presented in the Appendix\footnote{Note that the integrals $F_{11}^{(0)}$, $F_{13}^{(0)}$, and $F_{14}^{(0)}$ do not enter the cross section and presented only for completeness.}. 

%
 
 
\begin{figure}
	\centering\includegraphics[width=0.6\linewidth]{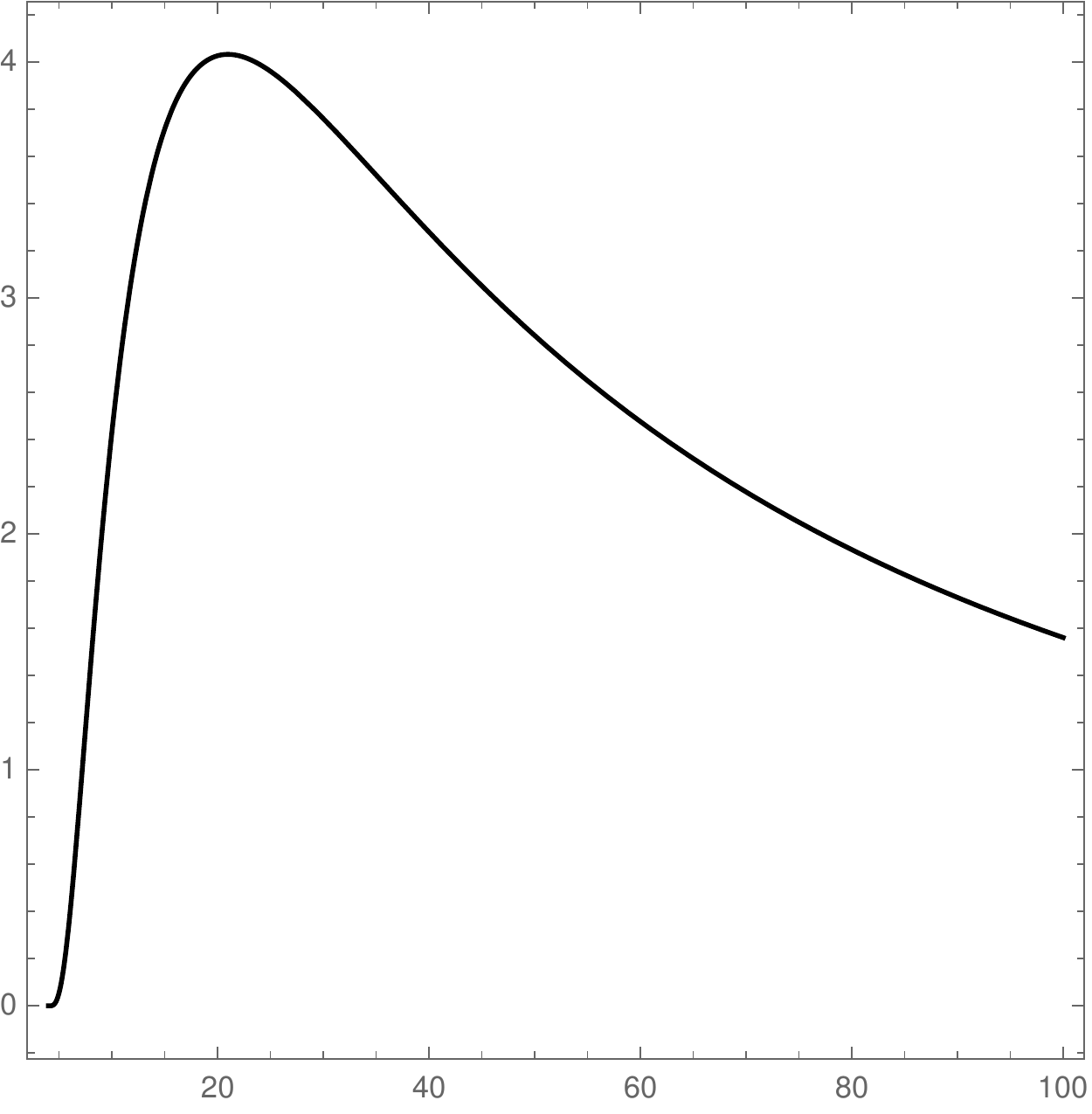}
	\setlength{\unitlength}{\linewidth}
	\begin{picture}(0,0)
	\put(-0.65,0.3){\rotatebox{90}{$\tilde{\sigma}$}}
	\put(-0.3,-0.02){$\sqrt{s}$}
	\end{picture}
	\caption{The normalized cross-section $ \tilde{\sigma}= \sigma/\left[\left(\frac{\alpha}{\pi}\right)^2\left(\frac{\alpha_s}{4\pi}\right)^2 e_Q^2\right]$ as a function of $\sqrt{s}$.}
	\label{fig:cross-section}
\end{figure}
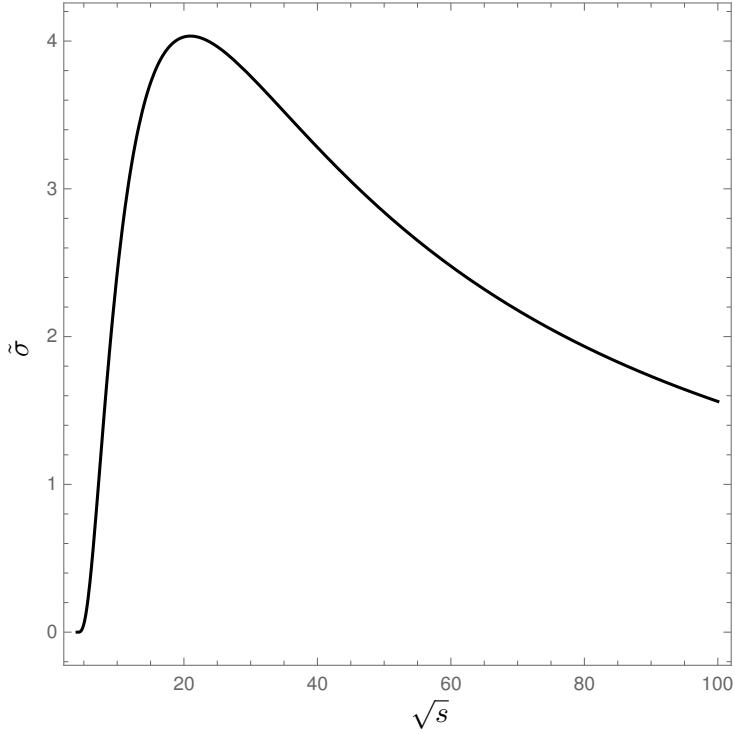

To check the obtained result we have performed the same calculation of total cross-section by squaring tree level matrix element and integrating it numerically over the final particles phase space with the use of massive Monte Carlo algorithm \texttt{RAMBO} \cite{rambo}. The comparison between analytical and numerical\footnote{We used non-adaptive Monte Carlo with 5 million sampling points.} results for the normalized cross-section\footnote{The correct relation of $\sigma$ and $\tilde{\sigma}$ with proper account of $m_Q$ reads $\sigma(s) = \frac{1}{m_Q^2}\left(\frac{\alpha}{\pi}\right)^2\left(\frac{\alpha_s}{4\pi}\right)^2 e_Q^2 \tilde{\sigma}(s/m_Q^2)$.} $\tilde\sigma(s) =  \sigma(s)/\left[\left(\frac{\alpha}{\pi}\right)^2\left(\frac{\alpha_s}{4\pi}\right)^2 e_Q^2\right]$ can be found in Table \ref{table:comparison}. The plot of normalized cross-section $\tilde{\sigma}$ is shown in Fig. \ref{fig:cross-section}. 

\begin{table}
\begin{center}
\begin{tabular}{ |c|c|c|c|c|c| } 
\hline
$\sqrt{s}$	 & $\tilde{\sigma}_{\text{exact}}$ & $\tilde{\sigma}_{\text{MC}}$ & $\sqrt{s}$	 & $\tilde{\sigma}_{\text{exact}}$ & $\tilde{\sigma}_{\text{MC}}$ \\ 
\hline
$5$ & $0.0544$ & $0.0544(1)$  & $60$ & $2.4767$ & $2.4789(70)$ \\
$10$ & $2.4303$ & $2.4309(6)$ & $70$ & $2.1785$ & $2.1739(70)$ \\
$15$ & $3.7172$ & $3.7165(20)$ & $80$ & $1.9338$ & $1.9338(70)$ \\
$20$ & $4.0277$ & $4.0277(30)$ & $100$ & $1.5614$ & $1.5642(80)$ \\
$30$ & $3.7620$ & $3.7605(40)$ & $200$ & $0.7304$ & $0.74(1)$ \\
$40$ & $3.2806$ & $3.2810(50)$ & $500$ & $0.2281$ & $0.22(1)$\\
 $50$ & $2.8415$ & $2.8374(60)$ & $1000$ & $0.0870$ & $0.09(1)$\\
\hline
\end{tabular}		
\end{center}
\caption{Comparison between analytical $\tilde{\sigma}_{\text{exact}}$ and numerical $\tilde{\sigma}_{\text{MC}}$ results for the normalized cross-section $\tilde\sigma =  \sigma/\left[\left(\frac{\alpha}{\pi}\right)^2\left(\frac{\alpha_s}{4\pi}\right)^2 e_Q^2\right]$ at different values of $\sqrt{s}$.}
\label{table:comparison}
\end{table}

\section{Conclusion}

In the present paper we have elaborated a method to treat the multiloop integrals in the case when they can not be expressed via polylogarithmic functions. Our method is based on the notion of $\epsilon$-regular basis of master integrals defined as a set of master integrals which are finite and linearly independent at $\epsilon=0$. This basis exists in any multiloop setup. The advantage of using this basis is that it allows one to calculate any finite physical quantity circumventing the necessity to expand the master integrals in $\epsilon$. This is especially advantageous for the non-polylogarithmic cases, when expansion in $\epsilon$ is typically quite complicated. Our approach results in the iterated integrals with almost all weights being rational functions. As an illustration of the advocated technique, we have calculated the photon contribution to the total Born cross section of the process $e^+e^-\to 2(Q\bar Q)$. We have expressed this cross section via iterated integrals with only the right-most integration weight being transcendental function. Alternatively, our result can be presented as a one-fold integral of the expression depending on dilogarithms, complete elliptic integrals, and elementary functions. We anticipate our method to be applicable to other problems where the non-polylogarithmic integrals are involved.

\acknowledgments
This work is supported by the grants of the ``Basis'' foundation for theoretical physics and mathematics.
 
\appendix
\section{$F_1^{(0)},\ldots, F_{14}^{(0)}$ via iterated integrals.}

Let us present here the explicit formulas for the leading terms  $F_1^{(0)},\ldots, F_{14}^{(0)}$.
The integrals $F_1^{(0)}$, $F_2^{(0)}$, and $F_3^{(0)}$ are expressed via $j_1|_{d=2}$, Eq. \eqref{eq:j1[d=2]}. We have

\begin{align}
F_1^{(0)}(s) =& (s-16) f(s) = \frac{16\pi (s-16)}{s}\left[
\mathrm{K}(1-k_{-})\mathrm{K}(k_{+}) - \mathrm{K}(k_{-})\mathrm{K}(1-k_{+})
\right] ,\label{eq:F10}\\
F_2^{(0)}(s) =& \frac{3s}{2(s-16)}F_1^{(0)}(s) - s\frac{d}{ds}F_1^{(0)}(s) , \\ 
F_3^{(0)}(s) =& \frac{12s}{(s-16)^2}F_1^{(0)}(s) + \frac{s(s-64)}{4(s-16)}\frac{d}{ds}F_1^{(0)}(s)  + \frac{s^2}{2}\frac{d^2}{ds^2}F_1^{(0)}(s) , \\
F_4^{(0)}(s) =& \frac{3}{2} I_1\left[
\frac{\beta}{s-16}
\right] - I_2\left[
\frac{1}{s\beta}
\right],\\
F_5^{(0)}(s) =& I_1\left[
\frac{1}{s}
\right] , \\
F_6^{(0)}(s) =& -\frac{3}{2}I_1\left[
\frac{1}{s\beta^2}, \frac{\beta}{s-16}
\right] + I_2\left[
\frac{1}{s\beta^2}, \frac{1}{s\beta}
\right] , \\
F_7^{(0)}(s) =& \frac{3}{2}I_1\left[
\frac{1}{s\beta}, \frac{\beta}{s-16}
\right] - I_2\left[
\frac{1}{s\beta}, \frac{1}{s\beta}
\right] , \\
F_8^{(0)}(s) =& \frac{1}{2}I_1\left[
\frac{1}{s\beta}
\right] - \frac{3}{2}I_1\left[
\frac{1}{s},\frac{\beta}{s-16}
\right] + I_2\left[
\frac{1}{s},\frac{1}{s\beta}
\right] ,\\
F_9^{(0)}(s) =& -\frac{1}{2}I_1\left[
\frac{1}{s\beta}, \frac{1}{s\beta}
\right] - \frac{3}{2}I_1\left[
\frac{1}{s\beta}, \frac{1}{s\beta^2}, \frac{\beta}{s-16}
\right] + I_2\left[
\frac{1}{s\beta}, \frac{1}{s\beta^2}, \frac{1}{s\beta}
\right] \nonumber \\
& + \frac{3}{2}I_1\left[
\frac{1}{s\beta}, \frac{1}{s}, \frac{\beta}{s-16}
\right] - I_2\left[
\frac{1}{s\beta}, \frac{1}{s}, \frac{1}{s\beta}
\right] , \\
F_{10}^{(0)}(s) =& \frac{3}{2}I_1\left[
\frac{1}{s}, \frac{1}{s\beta}, \frac{\beta}{s-16}
\right] - I_2\left[
\frac{1}{s}, \frac{1}{s\beta}, \frac{1}{s\beta}
\right] , \\
F_{11}^{(0)}(s) =& -\frac{1}{2}I_1\left[
\frac{1}{s\beta^2}, \frac{1}{s\beta}
\right] + \frac{1}{2}I_1\left[
\frac{1}{s}, \frac{1}{s\beta}
\right] - \frac{3}{4}I_1\left[
\frac{1}{s\beta^2}, \frac{1}{s\beta^2}, \frac{\beta}{s-16}
\right] \nonumber \\
& +\frac{1}{4}I_2\left[
\frac{1}{s\beta^2}, \frac{1}{s\beta^2}, \frac{1}{s\beta}
\right] + \frac{3}{2}I_1\left[
\frac{1}{s\beta^2}, \frac{1}{s}, \frac{\beta}{s-16}
\right] - I_2\left[
\frac{1}{s\beta^2}, \frac{1}{s}, \frac{1}{s\beta}
\right] \nonumber \\
& +\frac{3}{8}I_1\left[
\frac{1}{s},\frac{1}{s\beta^2},\frac{\beta}{s-16}
\right] - \frac{1}{4}I_2\left[
\frac{1}{s},\frac{1}{s\beta^2},\frac{1}{s\beta}
\right] - \frac{3}{2}I_1\left[
\frac{1}{s},\frac{1}{s},\frac{\beta}{s-16}
\right] \nonumber \\
& + I_2\left[
\frac{1}{s}, \frac{1}{s}, \frac{1}{s\beta}
\right] - \frac{9}{8}I_1\left[
\frac{1}{s\beta},\frac{1}{s\beta},\frac{\beta}{s-16}
\right] + \frac{3}{4}I_2\left[
\frac{1}{s\beta},\frac{1}{s\beta},\frac{1}{s\beta}
\right] , \\
F_{12}^{(0)}(s) =& -\frac{1}{4}I_1\left[
\frac{1}{s\beta}, \frac{1}{s}
\right] + \frac{3}{2}I_1\left[
\frac{1}{s\beta}, \frac{1}{s\beta}, \frac{\beta}{s-16}
\right] - I_2\left[
\frac{1}{s\beta}, \frac{1}{s\beta}, \frac{1}{s\beta}
\right] , \\
F_{13}^{(0)}(s) =& -\frac{1}{2}I_1\left[
\frac{1}{s\beta},\frac{1}{s\beta^2},\frac{1}{s\beta}
\right] + \frac{1}{2}I_1\left[
\frac{1}{s\beta},\frac{1}{s},\frac{1}{s\beta}
\right] - \frac{1}{4}I_1\left[
\frac{1}{s\beta},\frac{1}{s\beta},\frac{1}{s}
\right] \nonumber \\
& -\frac{3}{8}I_1\left[
\frac{1}{s\beta},\frac{1}{s\beta^2},\frac{1}{s\beta^2},\frac{\beta}{s-16}
\right] + \frac{1}{4}I_2\left[
\frac{1}{s\beta},\frac{1}{s\beta^2},\frac{1}{s\beta^2},\frac{1}{s\beta} 
\right] \nonumber \\
& + \frac{3}{2}I_1\left[
\frac{1}{s\beta},\frac{1}{s\beta^2},\frac{1}{s},\frac{\beta}{s-16}
\right] - I_2\left[
\frac{1}{s\beta}, \frac{1}{s\beta^2}, \frac{1}{s}, \frac{1}{s\beta}
\right] \nonumber \\
& +\frac{3}{8}I_1\left[
\frac{1}{s\beta},\frac{1}{s},\frac{1}{s\beta^2},\frac{\beta}{s-16}
\right] - \frac{1}{4}I_2\left[
\frac{1}{s\beta},\frac{1}{s},\frac{1}{s\beta^2},\frac{1}{s\beta}
\right] \nonumber \\
& -\frac{3}{2}I_1\left[
\frac{1}{s\beta},\frac{1}{s},\frac{1}{s},\frac{\beta}{s-16}
\right] + I_2\left[
\frac{1}{s\beta},\frac{1}{s},\frac{1}{s},\frac{1}{s\beta}
\right] \nonumber \\ 
& + \frac{3}{8}I_1\left[
\frac{1}{s\beta}, \frac{1}{s\beta}, \frac{1}{s\beta}, \frac{\beta}{s-16}
\right] - \frac{1}{4}I_2\left[
\frac{1}{s\beta},\frac{1}{s\beta}, \frac{1}{s\beta}, \frac{1}{s\beta}
\right] , \\
F_{14}^{(0)}(s) =& -\frac{1}{2}I_1\left[
\frac{1}{s}, \frac{1}{s\beta}, \frac{1}{s\beta}
\right] - \frac{1}{4}I_1\left[
\frac{1}{s\beta}, \frac{1}{s\beta}, \frac{1}{s}
\right] - \frac{3}{2}I_1\left[
\frac{1}{s},\frac{1}{s\beta},\frac{1}{s\beta^2},\frac{\beta}{s-16}
\right] \nonumber \\
& + I_2\left[
\frac{1}{s}, \frac{1}{s\beta}, \frac{1}{s\beta^2}, \frac{1}{s\beta}
\right] + \frac{3}{2}I_1\left[
\frac{1}{s},\frac{1}{s\beta}, \frac{1}{s}, \frac{\beta}{s-16}
\right] - I_2\left[
\frac{1}{s}, \frac{1}{s\beta}, \frac{1}{s}, \frac{1}{s\beta}
\right] \nonumber \\
& + \frac{3}{2}I_1\left[
\frac{1}{s\beta}, \frac{1}{s\beta}, \frac{1}{s\beta}, \frac{\beta}{s-16}
\right] - I_2\left[
\frac{1}{s\beta}, \frac{1}{s\beta}, \frac{1}{s\beta} , \frac{1}{s\beta} 
\right] .
\end{align}
where we introduced the following notation for iterated integrals:
\begin{align}
I\left[
w_1(s),\ldots , w_n(s)
\right] &= \int_{16}^sds_1 w_1(s_1)\int_{16}^{s_1}ds_2 w_2(s_2)\ldots \int_{16}^{s_{n-1}}ds_n w_n(s_n) , \\
I_{1,2} \left[
w_1(s),\ldots , w_n(s)
\right] &= I \left[
w_1(s),\ldots , w_n(s) F_{1,2}^{(0)}(s)
\right].
\end{align}
It is remarkable, that all weights, apart from the right-most ones, are restricted to the three-letter alphabet
$\big\{\frac{1}{s},\  \frac{1}{s\beta}=\frac{1}{\sqrt{s(s-4)}},\ \frac{1}{s\beta^2}=\frac{1}{s-4}\big\}$.

Note that the iterated integrals above can be easily turned into one-fold integrals of multiple polylogarithms and complete elliptic integrals. Indeed,
\begin{multline}
	I\left[	w_1(s),\ldots , w_n(s)\right] =\int_{16}^{s}ds_n w_n(s_n)  \left[\int_{s_{n}}^sds_{n-1} w_{n-1}(s_{n-1})\ldots \int_{s_2}^{s}ds_1 w_1(s_1)\right]\,,
\end{multline}
and, since the weights $w_1,\ldots,w_{n-1}$ are rational, the bracketed quantity is expressed via polylogarithms.
This may be convenient for numerical purposes. 
We present the corresponding expressions for the integrals $F_{4-10}^{(0)}$ and $F_{12}^{(0)}$ entering the cross section:
\begin{align}
F_4^{(0)}(s)&=\frac{\beta s  F_1^{(0)}(s)}{s-4}-\intop_{16}^s ds_1\frac{4 \beta_1 (s_1+2) F_1^{(0)}(s_1)}{(s_1-16) (s_1-4)^2},\\
F_5^{(0)}(s)&=\intop_{16}^s ds_1\frac{F_1^{(0)}(s_1)}{s_1},\\
F_{6}^{(0)}(s)&=\intop_{16}^{s}ds_1 \frac{\beta_1 F_1^{(0)}(s_1) }{
	(s_1-4)^2}\bigg\{-s_1-\frac{4   (s_1+2) }{s_1-16}\left[2
\ln \tfrac{\beta_1 }{\beta}+\ln
\tfrac{s_1}{s}\right]
\bigg\},\\
F_{7}^{(0)}(s)&=\intop_{16}^{s}ds_1 \frac{ F_1^{(0)}(s_1) }{
	(s_1-4)^2}\bigg\{	s_1-4+
	\frac{4 \beta_1  (s_1+2)}{s_1-16} \left[\ln \tfrac{1+\beta_1}{1-\beta_1 }-\ln\tfrac{1+\beta}{1-\beta}\right]
	\bigg\},\\
F_{8}^{(0)}(s)&=\intop_{16}^{s}ds_1 \frac{\beta_1 F_1^{(0)}(s_1) }{2
		(s_1-4)^2}\bigg\{-s_1+4-\frac{8 \beta_1  (s_1+2) \ln
		\frac{s_1}{s}}{s_1-16}
	\bigg\},\\
F_{9}^{(0)}(s)&=\intop_{16}^{s}ds_1 \frac{\beta_1 F_1^{(0)}(s_1) }{2
		(s_1-4)^2}\bigg\{
	 (s_1+4)\left[\ln \tfrac{1+\beta_1
	}{1-\beta_1 }-\ln
\tfrac{1+\beta}{1-\beta}\right]\nonumber\\
	&-\frac{16(s_1+2)}{s_1-16} \left[\ln
	\tfrac{1+\beta}{1-\beta} \ \ln \tfrac{\beta_1
}{\beta}+\text{Li}_2(-\beta_1 )-\text{Li}_2(\beta_1
)-\text{Li}_2(-\beta)+\text{Li}_2(\beta)\right]
	\bigg\},\\
	F_{10}^{(0)}(s)&=\intop_{16}^{s}ds_1 \frac{F_1^{(0)}(s_1) }{
	(s_1-4)^2}\bigg\{
-(s_1-4)\ln \tfrac{s_1}{s}
	-\frac{\beta_1  (s_1+2)}{s_1-16} \bigg[\tfrac12\ln
\tfrac{1+\beta_1 }{1-\beta_1 } \ln s_1+\tfrac12\ln
\tfrac{1+\beta}{1-\beta} \ln s\nonumber\\
&
-\ln\tfrac{1+\beta_1 }{1-\beta_1 } \ln s-\text{Li}_2\left(\tfrac{1-\beta_1 }{2}\right)+\text{Li}_2\left(\tfrac{1+\beta_1 }{2}\right)
+\text{Li}_2\left(\tfrac{1-\beta}{2}\right)-\text{Li}_2\left(\tfrac{1+\beta}{2}\right)
\bigg]
	\bigg\},\\
F_{12}^{(0)}(s)&=\intop_{16}^{s}ds_1 \frac{F_1^{(0)}(s_1) }{
	(s_1-4)^2}\bigg\{\frac{(3 s_1+4)(s_1-4) }{4  s_1}\left[\ln
\tfrac{1+\beta}{1-\beta}-\ln \tfrac{1+\beta_1
}{1-\beta_1 }\right]\nonumber\\
&
-\frac{2 \beta_1  (s_1+2)}{s_1-16}\left[\ln
	\tfrac{1+\beta}{1-\beta}-\ln \tfrac{1+\beta_1
	}{1-\beta_1 }\right]^2
	\bigg\}.
\end{align}
Here $\beta_1=\sqrt{1-4/s_1}$ and $ F_1^{(0)}(s)$ is defined in Eq. \eqref{eq:F10} via complete elliptic integral $\mathrm{K}$.
%
%
%
%
%
%
%


\bibliographystyle{JHEP}
\providecommand{\href}[2]{#2}\begingroup\raggedright\endgroup

\end{document}